\documentclass[preprint,aps]{revtex4}
\newcommand{\be}{\begin{equation}} \newcommand{\ee}{\end{equation}} 
\newcommand{\bea}{\begin{eqnarray}}\newcommand{\eea}{\end{eqnarray}}

\usepackage{graphicx}
\begin{document}
\preprint {SINP/TNP/03-37}
\title{On the Real Spectra of Calogero Model with Complex Coupling}
\author{Pijush K. Ghosh}
\email{pijush@theory.saha.ernet.in}
\author{Kumar S. Gupta}
\email{gupta@theory.saha.ernet.in}
\affiliation{Theory Division,\\ Saha Institute of Nuclear Physics,\\ 
Kolkata 700 064, India.\\}
\begin{abstract} 
We study the eigenvalue problem of the rational Calogero model with the
coupling of the inverse-square interaction as a complex number. We show
that although this model is manifestly non-invariant under
the combined parity and time-reversal symmetry ${\cal{PT}}$, the eigenstates 
corresponding to the zero value of the generalized angular momentum have real
energies.
\end{abstract}
\maketitle

\newpage

The standard practice in quantum mechanics is to consider self-adjoint operators
so that the corresponding spectrum is real and consequently, the time-evolution
of the states is unitary. However, it has been found recently that a
non-Hermitian Hamiltonian invariant under the combined ${\cal{PT}}$ symmetry
can have real spectrum \cite{pt1,pt2,pt3,ali,ddt,ex1,ex2,ex3,bm,zn}. The
spectrum is
entirely real if the ${\cal{PT}}$ symmetry is unbroken. On the other hand,
part of the spectrum is real if the ${\cal{PT}}$ symmetry is broken
spontaneously\cite{pt1}.

In this Letter we study the rational Calogero model \cite{calo3,pr}
with complex coupling for the many-body interaction. 
Several extensions of the Calogero model with non-Hermitian Hamiltonians
have been studied in the literature \cite{bm,zn}. However, all these models
respect ${\cal {PT}}$ symmetry and lead to real spectrum. The Hamiltonian we
study is identical to that of the original $N$-body rational Calogero model
except that the coefficient of the inverse square interaction is complex.
This model is non-Hermitian, manifestly non-invariant under ${\cal {PT}}$, 
but nevertheless is shown to admit a partly real spectrum, with the ground
state energy always real.

Following the analysis of Calogero \cite{calo3}, the eigenvalue
equation of the rational Calogero model with complex coupling
for the many-body inverse-square interaction can be reduced to the eigenvalue
equation of an effective single particle Hamiltonian containing harmonic and 
inverse-square interaction. We show that for certain values of the complex
coupling of the inverse square interaction, 
this effective Hamiltonian can indeed be Hermitian when the generalized
angular momentum \cite{calo3} is zero. This however
happens only for the strongly attractive values of the inverse square
coupling term in the effective Hamiltonian. Such a strongly attractive 
single particle system in absence of the
harmonic term has already been analyzed by Case \cite{case}, who obtained the
corresponding spectrum in terms of an undetermined parameter which 
has the physical interpretation of
a cutoff in the coordinate space. In our analysis, we explicitly introduce a
cutoff which prevents the particles to occupy the same position
simultaneously, i.e. it serves as a cutoff in the coordinate space. 
In the analysis of ref. \cite{case}, the spectrum of
the strongly attractive model in absence of the harmonic term is unbounded 
from below. In the model under
consideration here, which includes the harmonic term, 
the spectrum is however bounded from below for finite
nonzero values of the cutoff.

The Hamiltonian of the rational Calogero model is given by
\be
H = - \sum^{N}_{i=1} \frac{{\partial}^2}{\partial x_i^2} +
\sum_{i \neq j} \left [ \frac{a^2 - \frac{1}{4}}{(x_i - x_j)^2} +
\frac{\Omega^2}{16} (x_i - x_j)^2 \right ]
\label{e0}
\ee
where $a$ and $\Omega$ are constants,
$x_i$ is the coordinate of the $i^{\rm th}$ particle and
units have been chosen such that $2 m {\hbar}^{- 2} = 1$.
The constant $a$ is conventionally taken to be real, which is required for
the operator $H$ to be Hermitian. In this Letter we shall however show that
for certain complex values of $a$, the eigenvalue problem
\be
H \Psi = E \Psi
\label{e1}
\ee
still admits a real spectrum with normalizable solutions. In the discussion
below, we therefore take $a = a_R + i a_I$ where $a_R$ and $a_I$ are the
real and imaginary parts of $a$. The Hamiltonian $H$ is invariant under
parity $x_i \rightarrow - x_i$, while non-invariant under the time-reversal
symmetry $ i \rightarrow -i$. Thus, the combined ${\cal {PT}}$
 symmetry is not respected by $H$. 

Following \cite{calo3}, we consider the above eigenvalue equation
in a sector of configuration
space corresponding to a definite ordering of particles given by
$x_1 \geq x_2 \geq
\cdots \geq x_N$. The translation-invariant
 eigenfunctions of the Hamiltonian $H$ can be written as
\be
\Psi = \prod_{i <j} \left (x_i - x_j \right )^{a + \frac{1}{2}} \
\phi (r) \ P_k (x),
\label{e2}
\ee
where $x \equiv (x_1, x_2, \dots, x_N)$,
\be
r^2 = \frac{1}{N} \sum_{i < j} (x_i - x_j)^2 \ \
\label{e3}
\ee
and $P_k (x)$ is a translation-invariant as well as  homogeneous
polynomial of degree $k(\geq 0)$ which satisfies the
equation
\be
\left[ \sum^{N}_{i=1}\frac{{\partial}^2}{\partial x_i^2}
+ \sum_{i \neq j} \frac{ 2 (a + \frac{1}{2}) }{(x_i -
x_j)}  \frac{{\partial}}{\partial x_i}
\right] P_k (x) = 0.
\label{e4}
\ee
The existence of
complete solutions of (\ref{e4}) for real $a$ has been discussed by Calogero
\cite{calo3}. For both real and complex $a$, $P_0(x)$ is a
constant and is a solution of Eqn. (5). 
For $k \neq 0$, the solutions of Eqn. (\ref{e4}) 
for complex $a$ are obtained by analytic continuation in the parameter 
space $a$. 

Substituting Eqn. (\ref{e2}) in Eqn. (\ref{e1}), using Eqns.
(\ref{e3}- \ref{e4}) and making a further substitution
$\phi= r^{-(\mu +\frac{1}{2})} \Phi$ we get
\be
\tilde{H} \Phi = E \Phi,
\label{e5}
\ee
where
\be
\tilde{H} =  - \frac{d^2}{dr^2} + \frac{(\mu^2 - \frac{1}{4} )}{r^2}
+ \omega^2 r^2 ,
\ee
with $ \omega^2 = \frac{1}{8}  \Omega^2 N $ and 
\be
\mu = k + \frac{1}{2}(N - 3) + \frac{1}{2} N (N-1)(a_R + i a_I +
\frac{1}{2}). 
\ee
The operator $\tilde{H}$ can be interpreted as the
effective Hamiltonian of a particle
in the combined harmonic plus inverse-square interaction.
Let us now choose the real part of $a$ as 
\be
a_R = -\frac{N-3}{N (N-1)} - \frac{1}{2},
\ee
and keep $a_I$ as an arbitrary real number. 
It may be noted that the wavefunction $\Psi$ is square-integrable for
$a_R+\frac{1}{2} > -\frac{1}{2}$ \cite{us1,we}, which is indeed satisfied
for the above choice of $a_R$ for $N \geq 3$. With the above choice of the 
complex coupling $a$, for $k=0$ we see that 
\be
\mu = \frac{i}{2} N (N-1) a_I \equiv i \nu
\ee
is purely imaginary and the operator $\tilde{H}$ takes the form
\be
\tilde{H} = - \frac{d^2}{dr^2} - \frac{(\nu^2 + \frac{1}{4} )}{r^2}
+ \omega^2 r^2 ,
\label{eh}
\ee
\noindent 
which is Hermitian and is expected to have real eigenvalues. However, for the 
same choice of $a_R$, when $k \neq 0$, $\mu$ is in general a complex
quantity. Thus the eigenvalues of $\tilde{H}$ in the $k \neq 0$ sector in
general would not be real. Below we shall first find the eigenvalues of the 
operator $\tilde{H}$  for $k=0$ and then provide a suitable interpretation
of the states in the $k \neq 0$ sector.


We now proceed to solve the eigenvalue problem given in Eqn. (\ref{e5}) with 
$\tilde{H}$ given by Eqn. (\ref{eh}). 
In this case,
the inverse-square interaction is necessarily in the strongly attractive
regime for $\nu \neq 0$.
As mentioned before, we put a cutoff 
$r_0$ in the coordinate space which leads to the boundary condition 
$\Phi (r) = 0$ for $r = r_0$. We also demand that $\Phi(r) \in {\rm
L}^2(R^+)$.
Defining $q=\omega r^2$ and $\Phi=q^{-\frac{1}{4}} \chi(q)$,
the eigenvalue equation $\tilde{H} \Phi = E \Phi$ can be written as,
\be
\frac{d^2 \chi}{d q^2} + \left [ -\frac{1}{4} + \frac{1}{4 q^2} ( 1 + \nu^2)
+ \frac{E}{4 w q} \right ] \chi =0,~~~~\chi (q_0 = w r^2_0) = 0.
\label{eq1}
\ee
\noindent The above equation can be identified as the Whittaker's equation,
the two linearly independent solutions of which are given by
\be
W_{\frac{E}{4 \omega}, \pm \frac{  i \nu}{2}} (q) = 
e^{-\frac{q}{2}} q^{\frac{1 \pm i  \nu}{2}} \
M\left (\frac{1 \pm i  \nu}{2}-\frac{E}{4 \omega}, 1 \pm i  \nu, q \right ),
\label{eq1.1}
\ee
\noindent where $ M\left (\frac{1 \pm i \nu}{2}-\frac{E}{4 \omega}, 1 \pm
i \nu, q \right )$ is Kummer's function \cite{abr}. 
The general solution
of Eqn. (\ref{eq1}) satisfying the boundary condition at $q_0$ is given by
\be
\chi(q) = A \left [ W_{\frac{E}{4 \omega}, \frac{i \nu}{2}} (q) \
W_{\frac{E}{4 \omega}, -\frac{i \nu}{2}} (q_0) -
W _{\frac{E}{4 \omega}, - \frac{i \nu}{2}} (q) \ 
W_{\frac{E}{4 \omega}, \frac{i \nu}{2}} (q_0) \right ],
\label{eq1.2}
\ee
\noindent where $A$ is the normalization constant. 
In the limit $q \rightarrow \infty$, we have \cite{abr}
\be
\frac{M(a,b,q)}{\Gamma(b)} \equiv \frac{ e^{i \pi a} q^{-a}}{\Gamma(b-a)}
+ \frac{ e^q q^{a-b}}{\Gamma(a)} + O\left(\frac{1}{q} \right ).
\label{eq1.3}
\ee
Thus, as $q \rightarrow \infty$, 
\be
\chi (q) \rightarrow A e^{\frac{q}{2}} \left [ \frac{\Gamma(1+i \nu)}
{\Gamma(\frac{1+i \nu}{2}-\frac{E}{4 \omega})}  
W_{\frac{E}{4 \omega}, -\frac{i \nu}{2}} (q_0)
- \frac{\Gamma(1-i \nu)}{\Gamma(\frac{1-i \nu}{2}-\frac{E}{4 \omega})}
W_{\frac{E}{4 \omega}, \frac{i \nu}{2}} (q_0) \right ].
\ee
The square integrability of $\chi$ can thus be ensured if the quantity in
the parenthesis in Eqn. (16) is identically zero, i.e.
\be
\frac{W_{\frac{E}{4 \omega}, -\frac{i \nu}{2}} (q_0)}{
W_{\frac{E}{4 \omega}, \frac{i \nu}{2}} (q_0)}
 = \frac{\Gamma(1-i \nu)}{\Gamma(1+i \nu)}
\frac{\Gamma(\frac{1+i \nu}{2}-\frac{E}{4 \omega})}{\Gamma(\frac{1- i \nu}{2}
-\frac{E}{4 \omega})}.
\label{eq1.4}
\ee
We are interested in the solution of the eigenvalue equation in the limit of
a small cutoff, i.e. when $q_0 \rightarrow 0$. In this limit, we have 
$ W_{\frac{E}{4 \omega}, -\frac{i \nu}{2}} (q_0) \rightarrow
q_0^{\frac{1-i \nu}{2}}$ \cite{abr}. The energy eigenvalues are thus 
determined from the equation,
\be
q_0^{- i \nu} = \frac{\Gamma(1-i \nu)}{\Gamma(1+i \nu)}
\frac{\Gamma(\frac{1+i \nu}{2}-\frac{E}{4 \omega})}{\Gamma(\frac{1- i \nu}{2}
-\frac{E}{4 \omega})},
\label{eq1.5}
\ee
or equivalently from
\be
e^{-i ( \nu {\rm ln} q_0 + 2 \theta )} = e^{2 i \alpha },
\label{eq1.55}
\ee
where $\theta$ and $\alpha$ are the arguments in the polar representation 
of $\Gamma(1-i \nu)$ and 
$\Gamma(\frac{1+i \nu}{2}-\frac{E}{4 \omega})$ respectively. The spectrum is
obtained by solving Eqn. (19) graphically. 
We have studied Eqn. (19) using Mathematica for
different
values of $\nu$ and $q_0$. For a fixed $\nu$ and $q_0$, there is one
negative
energy bound state and infinitely many positive energy bound states.
Moreover,
the energy levels are not equispaced. The absolute value of the negative
energy depends on the choice of $q_0$. For smaller values for $q_0$, $|E|$
for the negative energy eigenstate increases. We plot our results in Fig.1 
and Fig. 2.

\begin{figure}  

\begin{minipage}[t]{0.45\linewidth}
\centering
\includegraphics[width=1.0\textwidth, height=0.25\textheight]{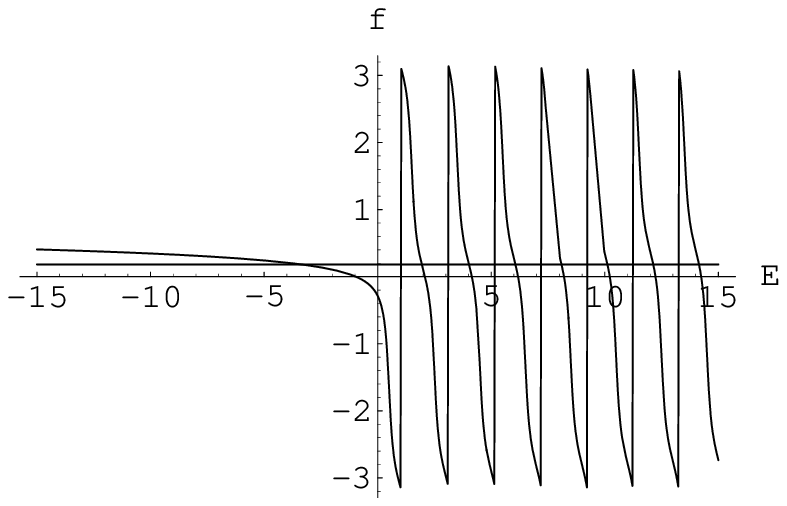}
\caption{{\small 
\label{fig2}     
A plot of Eqn. (\ref{eq1.55})
with $w = 0.25$, $\nu = 0.3 $ and
$q_0 =0 .1$. The horizontal straight line corresponds to the value of the
l.h.s of Eqn. (\ref{eq1.55}) and $f= 2 \alpha$.
}}
\end{minipage}%
\hfill
\begin{minipage}[t]{0.45\linewidth}
\centering   
\includegraphics[width=1.0\textwidth, height=0.25\textheight]{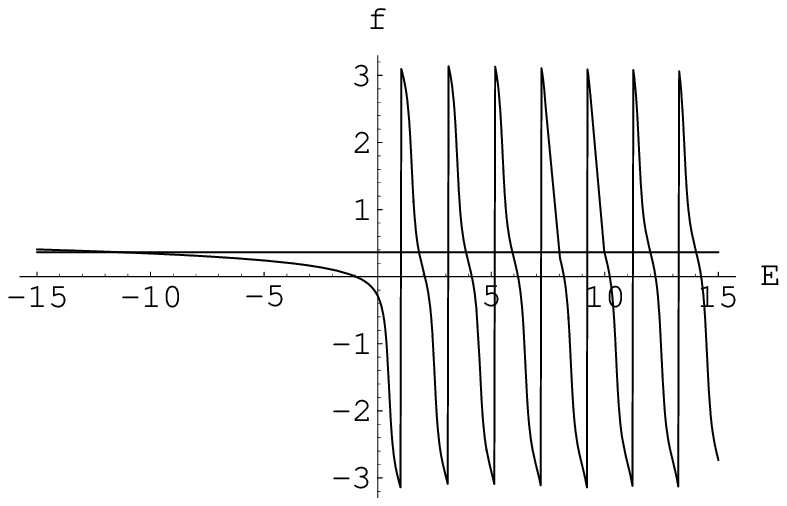}
\caption{{\small 
\label{fig1} A plot of Eqn. (\ref{eq1.55}) 
with $w = 0.25$, $\nu = 0.3 $ and
$q_0 = .03$. The negative energy state has a lower eigenvalue 
due to the choice of a smaller value of the cutoff $q_0$.
}}
\end{minipage}%
\end{figure}  

The following
points about the spectrum in the $k =0$ sector may be noted: \\
\noindent
1) As shown above, even for certain complex values of the inverse-square
coupling, the rational Calogero model admits a real spectrum in the $k=0$
sector. The real part of the relevant coupling is $N$ dependent and is 
given by Eqn. (9) whereas the imaginary part of the coupling is arbitrary.
Thus there is a range of values of the parameter $a$ for which the spectrum
in the $k=0$ sector is real.

\noindent
2) The
spectrum here for a given value of $q_0$ and $\nu$ is bounded from below. In 
order to see this, consider $E = - {\cal {E}} $ with ${\cal {E}} \rightarrow
\infty$. In this limit, the argument of the $\Gamma(\frac{1+i
\nu}{2}-\frac{E}{4 \omega})$ can be approximated by $ \frac{\nu}{2} \psi (
\frac{1}{2} + \frac{{\cal {E}}}{4 w} )$ and Eqn. (18) can be written as 
\be
-\nu ~ {\rm ln} q_0 - 2 \theta = \nu \psi \left ( \frac{1}{2} +
\frac{{\cal {E}}}{4 w} \right ),
\ee
where $\psi$ denotes the digamma function \cite{abr}. For positive values of
its argument, the digamma function is monotonically increasing. Thus, for
any given finite but nonzero value of $q_0$ and $\nu$, $E \rightarrow - 
\infty$ is not a part of the spectrum. 

\noindent
3) This system  generically admits a single negative energy eigenstate. 
Existence of
negative energy eigenstates in the Calogero model is associated with the   
self-adjoint extension of the corresponding Hamiltonian \cite{us1,we}. We
have however not studied the self-adjointness of the effective Hamiltonian 
${\tilde{H}}$ here.

	As mentioned before, the parameter $\mu$ is  complex when
$k \neq 0$. In that case the effective Hamiltonian would not be Hermitian 
and the states with $k \neq 0$ are expected to have
complex eigenvalues. These states with complex values of the energy would 
decay under time evolution. Thus
the spectrum obtained here has a part which is stable under time evolution 
and a decaying part corresponding to complex eigenvalues. This situation is 
similar to what happens in the case of many nuclei, 
where only a few energy states are 
stable and the others are resonances with characteristic decay widths. It 
would be interesting to find such physical applications of the model 
described here.


\acknowledgments{ The work of PKG is supported(DO No. SR/FTP/PS-06/2001) by
the SERC, DST, Govt. of India, under the Fast Track Scheme for Young
Scientists:2001-2002. }

\end{document}